\documentclass[twocolumn,showpacs,preprintnumbers,amsmath,amssymb,prb]{revtex4}

\usepackage{graphicx}
\usepackage{dcolumn}
\usepackage{bm}
\newcommand{\CuTeX} {{\mbox{Cu${}_2$Te${}_2$O${}_5$X${}_2$}}}
\newcommand{\CuTeCl} {{\mbox{Cu${}_2$Te${}_2$O${}_5$Cl${}_2$}}}
\newcommand{\CuTeBr} {{\mbox{Cu${}_2$Te${}_2$O${}_5$Br${}_2$}}}
\newcommand{\half}{{\ensuremath{\frac{1}{2}}}}

\begin{document}
\preprint{APS/123-QED}
\title{Incommensurate magnetic ordering in ${\CuTeX}$ (X=Cl, Br) 
studied by neutron diffraction}
\author{O. Zaharko,\cite{now} A. Daoud-Aladine, S. Streule, J. Mesot}
\affiliation{Laboratory for Neutron Scattering, ETHZ \& PSI, CH-5232  Villigen, Switzerland}
\author{P.-J. Brown}
\affiliation{Institut Laue-Langevin, 156X, 38042 Grenoble C\'{e}dex, France}
\author{H. Berger}
\affiliation{Institut de Physique de la Mati\`{e}re Complexe, EPFL,CH-1015 Lausanne, Switzerland}
\date{\today}

\begin{abstract}
We present the results of the first neutron powder and single 
crystal diffraction studies of the coupled spin tetrahedra systems 
${\CuTeX}$ (X=Cl, Br). Incommensurate antiferromagnetic order with 
the propagation vectors ${\bf{k}_{Cl}}\approx[0.150,0.422,\half]$,
${\bf{k}_{Br}}\approx[0.158,0.354,\half]$ sets in below  
$T_{N}$=18 K for X=Cl and 11 K for X=Br. 
No simple collinear antiferromagnetic or ferromagnetic arrangements
of moments within Cu${}^{2+}$ 
tetrahedra fit these observations. Fitting the diffraction data
to more complex but physically reasonable models with multiple helices
leads to a moment of 0.67(1)$\mu_B$/Cu${}^{2+}$ at 1.5~K for the Cl-compound. The reason 
for such a complex ground state may be 
geometrical frustration of the spins due to the intra- and 
inter-tetrahedral couplings having similar strengths. The magnetic moment in the 
Br- compound, calculated assuming it has the same magnetic structure as the Cl compound, is
only 0.51(5)$\mu_B$/Cu${}^{2+}$ at 1.5 K. In neither  compound has any 
evidence for a structural 
transition accompanying the magnetic ordering been found.
\end{abstract}

\pacs{75.10.Jm, 75.30.-m, 61.12.Ld}
\keywords{magnetic ordering, quantum spin system, neutron scattering}
\maketitle

Recently much experimental and theoretical effort has been invested 
in trying to understand
low-dimensional quantum spin systems.\cite{Diep03}
Reduced dimensionality and
frustration in such systems lead to interesting new ground states and 
spin dynamics.
It is known, for example, that one-dimensional (1D) dimerized or 
frustrated spin chains
have gapped singlet ground states. In two-dimensional (2D) spin systems, which
include the cuprate high-temperature superconductors, strong 
renormalization of the
spin excitations due to quantum fluctuations has been 
reported.\cite{Hayden91, Ronnow01}

\begin{figure}[tbh]
\caption {Temperature evolution of the lattice constants from  
high-resolution HRPT NPD data.
Circles denote the $a$ and squares the $c$ lattice constants.}
\includegraphics[width=86mm,keepaspectratio=true]{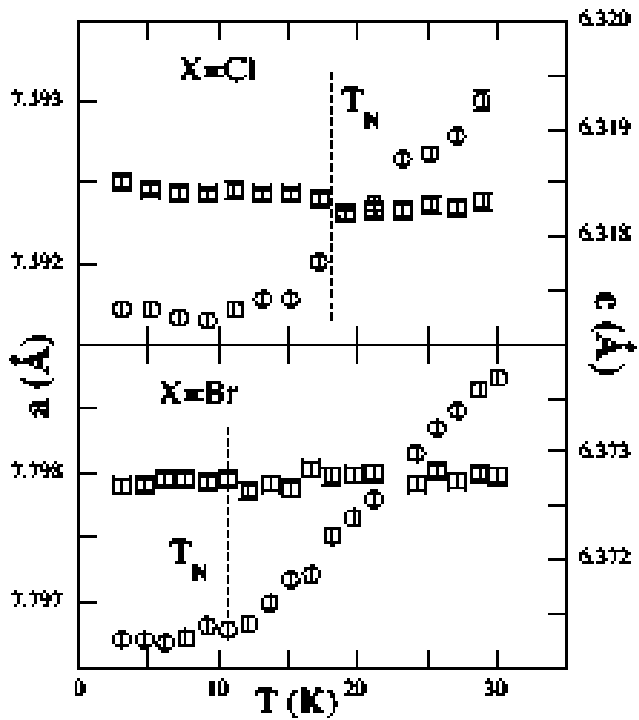}
\label{fig1}
             \end{figure}

\begin{figure}[tbh]
\caption {DMC neutron powder diffraction patterns of ${\CuTeCl}$. 
Arrows point to magnetic reflections.}
\includegraphics[width=86mm,keepaspectratio=true]{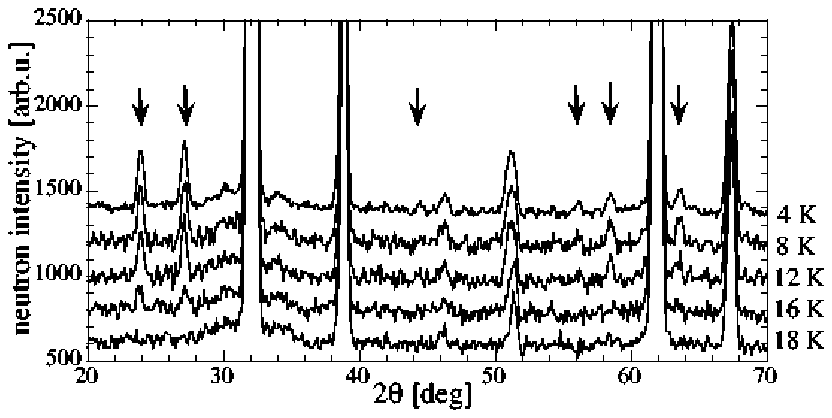}
\label{fig2}
             \end{figure}

\begin{figure}[tbh]
\caption {Temperature evolution of the lowest angle magnetic peak
($\approx\alpha, \beta, \half)$ of Cl (squares) and Br (circles)
compounds from DMC NPD data. Top: peak position in [deg.],
bottom: integrated intensities, normalized to the low temperature
value. The solid line represents the square of the S=$\half$ 
Brillouin function.}
\includegraphics[width=86mm,keepaspectratio=true]{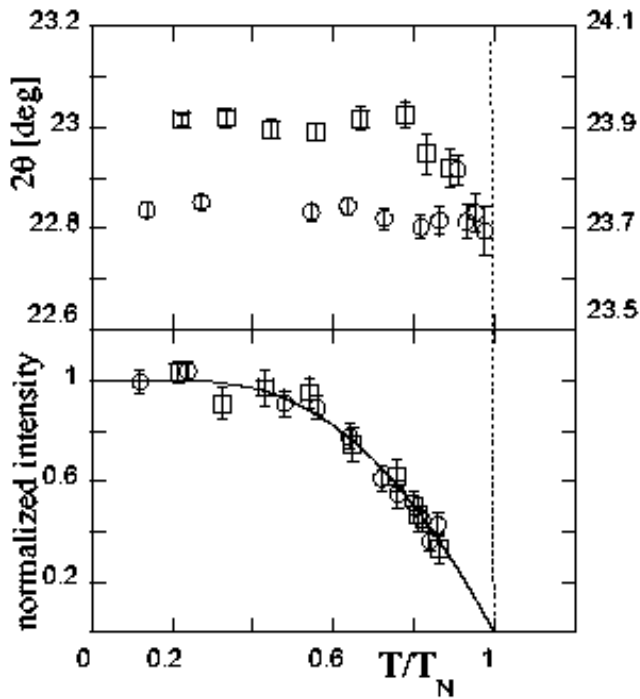}
\label{fig3}
\end{figure}

\begin{figure}[tbh]
\caption {a) The [001] projection of reciprocal space for the 
$\CuTeCl$ structure. Black points correspond to ${\bf{k}}$ and grey - to 
${\bf{k'}}$ magnetic reflections. b) Definition of the angles for the 
helical spin structure.}
\includegraphics[width=86mm,keepaspectratio=true]{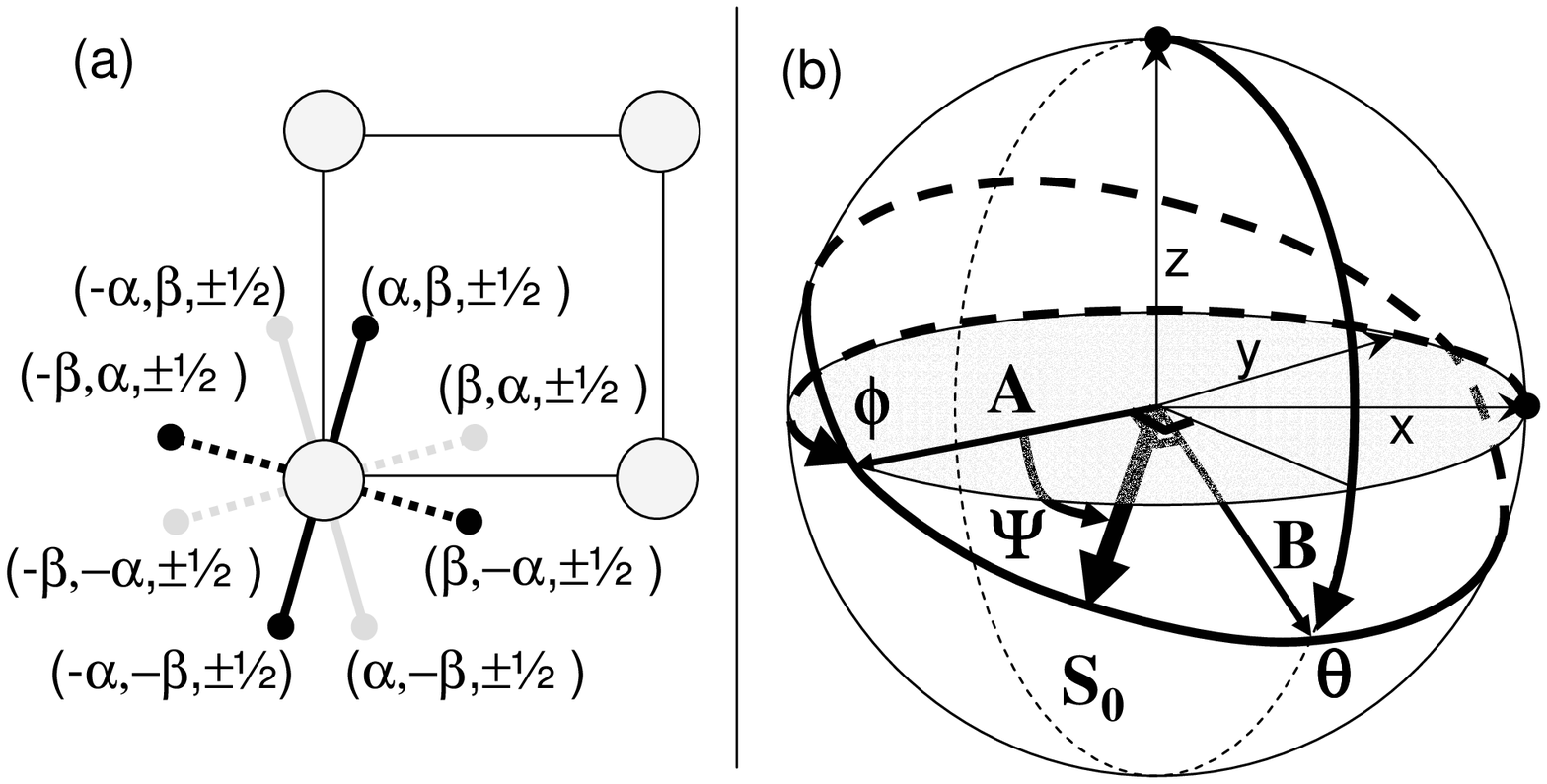}
\label{fig4}
             \end{figure}

The copper tellurates ${\CuTeX}$ (X=Cl, Br)\cite{Johnsson00} belong 
to a new family of such
compounds. They contain tetrahedral clusters of S=$\half$ Cu${}^{2+}$ 
spins aligned in
tubes along the $c$ direction and separated by lone pair cations in 
the $ab$ plane.
These systems are therefore ideally suited to study the interplay between the
built-in frustration within tetrahedra and magnetic coupling between them.
Both compounds crystallize in the non-centrosymmetric tetragonal 
space group $P\overline{4}$.
The four Cu${}^{2+}$ ions occupy a single set of general equivalent 
positions 4(h) ($x\approx 0.730$, $y\approx 0.453$, $z\approx 0.158$). 
The 4 ions clustered round the origin: Cu1 ($x, y, z$), Cu2 ($1-x, 1-y, z$), Cu3 
($y, 1-x, -z$) and Cu4 ($1-y, x, -z$) form an irregular 
tetrahedron with
two longer (Cu1-Cu2, Cu3-Cu4) and four shorter edges.\\
The magnetic susceptibility of both compounds shows a broad maximum 
around 20 K -- 30 K and drops rapidly
at lower temperatures,\cite{Johnsson00} as is typical of spin 
gapped systems. The strength of the coupling constant obtained
by fitting the susceptibility to a model in which the tetrahedra  
are isolated and all 4 inter-tetrahedral coupling constants have 
the same strength is 38.5~K and 43~K for X= Cl and Br, respectively.
Further magnetic susceptibility and specific heat 
measurements\cite{Lemmens01} indicate the onset of
antiferromagnetic (AF) order in the Cl-compound at $T_{N}$=18.2 K 
showing that the inter-tetrahedral coupling is substantial.
Interestingly,  the magnetic susceptibility is almost isotropic and thermal 
conductivity studies suggest strong spin-lattice coupling near 
$T_{N}$.\cite{Prester04} For the Br-compound an unusual phase 
transition
involving low-energy longitudinal magnetic modes around 11.4~K has 
been inferred from Raman scattering,\cite{Lemmens01} which was later 
attributed to the onset of magnetic order.\cite {Jensen03}\\
Raman light scattering experiments\cite{Gros03, Jensen03} in
${\CuTeX}$ compounds have been analyzed using a model in which 
the ion pairs Cu1-Cu2 and Cu3-Cu4 act as dimers within each
tetrahedron. The exchange constants $J_1$ and $J_2$ in this model 
are defined by the spin Hamiltonian
\begin{eqnarray}
H &=& J_{1} ({\bf{S}}_{1}\cdot {\bf{S}}_{3}+{\bf{S}}_{1}\cdot
{\bf{S}}_{4}+{\bf{S}}_{2}\cdot {\bf{S}}_{3}+{\bf{S}}_{2}\cdot
{\bf{S}}_{4} )
\nonumber\\
&+&
J_{2} ({\bf{S}}_{1}\cdot {\bf{S}}_{2}+{\bf{S}}_{3}\cdot {\bf{S}}_{4}).
\label{eq:eq2}
\end{eqnarray}
and were determined as:
${{J}_1}$=47.5 K and ${{J}_2}$/${{J}_1}$=0.7 for the Br-, and 
${{J}_1}$=40.7 K and ${{J}_2}$/${{J}_1}$=1 for the Cl-system. 
Many experimental and 
theoretical\ questions about  these interesting systems
still remain open. \cite{Brenig01, 
Brenig03, Valenti03, Totsuka02, Kotov04} 
Neutron elastic and inelastic scattering may provide 
answers to some of them by determining the magnetic ground states 
of the compounds and by probing their spin 
dynamics.
We report here the results of the first neutron diffraction 
experiments on the compounds. The experiments have been made on
powder and single crystal samples of ${\CuTeCl}$ and on
powder samples of 
${\CuTeBr}$. Both systems have been found to exhibit long-range 
incommensurate magnetic order.\\
High-purity powders of ${\CuTeX}$ and single crystals of 
${\CuTeCl}$ were prepared by the halogen vapor transport 
technique, using TeX${{}_4}$ and X${{}_2}$ as transport agents. 
Neutron powder diffraction (NPD) patterns were collected in the 
temperature range 1.5 K - 30 K, on the 
DMC instrument at SINQ, Switzerland, with a neutron wavelength of 
$\lambda$=4.2~\AA~(Cl) and $\lambda$=2.6~\AA~(Br) and on the 
high-resolution HRPT instrument at SINQ ($\lambda$=1.889~\AA) 
The neutron single crystal diffraction (NSCD) experiments on two 
crystals of dimensions 2.5 x 3 x15 mm${{}^3}$ and 2 x 3.5 x 6 
mm${{}^3}$ were carried out using the diffractometers TriCS at SINQ 
($\lambda$=1.18 \AA) and D15 ($\lambda$=1.17 \AA) at the high-flux 
ILL reactor, France.\\
The evolution of the lattice constants at low temperatures for 
${\CuTeX}$  X=Cl and Br is presented in Fig.~\ref{fig1}. The lattice 
contraction for both compounds is anisotropic above $T_{N}$: 
the change in the $ab$ plane being greater than that along $c$. Below $T_{N}$ 
the lattice constants change very little and no splitting or 
broadening of Bragg peaks, such as would occur in an accompanying 
structural transition, has been observed.\\
In ${\CuTeCl}$ below $T_{N}$=18~K magnetic peaks appeared 
(Fig.~\ref{fig2}), which cannot be indexed using simple multiples of the 
crystallographic unit cell, thus implying that the magnetic 
structure is incommensurate (ICM). The wave vector ${\bf{k}_{Cl}}
\approx[0.150,0.422,$\half$]$ was determined from the single 
crystal experiment.  The magnetic peaks in the ${\CuTeBr}$ NPD patterns 
appeared below $T_{N}$=11~K. Their topology is similar to that observed in the 
Cl-compound and corresponds to the wave vector 
${\bf{k}_{br}}\approx[0.158,0.354,$\half$]$.\\
The variation with temperature of the positions and intensities of the 
lowest angle magnetic peaks of both compounds is shown in Fig.~\ref{fig3}. 
The intensity varies as
the square of the S=$\half$ Brillouin function. 
The position of the ${\CuTeBr}$ peak is almost
constant and strictly incommensurate below $T_{N}$, whilst that of
${\CuTeCl}$ varies significantly in the vicinity of $T_{N}$.
The width of the magnetic peaks is resolution-limited, indicating the
long-range nature of the magnetic order.\\
The diffraction experiments on ${\CuTeCl}$ single crystals allowed us 
to clarify the magnetic symmetry and enumerate the magnetic domains.
The star of the wave vector   
${\bf{k}}$=$(\alpha,\beta,$\half$)$ in the space group $P\overline{4}$
has four arms shown by the black lines in
Fig.~\ref{fig4}a.  Each arm gives rise to a configuration domain; 
all have the same structure but possibly different populations.
In the crystal investigated on the 
TriCS instrument the only satellites observed  were
the pairs $H\pm{\bf{k}}$  around each
nuclear Bragg reflection $H$, so the 90 deg rotation domains are not 
present. In the crystal investigated on D15 not only were all four 
satellites around each $H$ observed, but in addition, 
another star of the wave vector 
${\bf{k'}}\approx[-0.15,0.42,$\half$]$ was present.
The ${\bf{k}}$ and ${\bf{k'}}$ vectors are not related by the 
symmetry elements of the group $P\overline{4}$, so the magnetic 
structures associated  with these two stars must be different. Pairs of vectors 
from the two stars may combine to generate a ${\bf{k}-\bf{k'}}$ structure. 
Careful comparison of
the integrated intensities in the ${\bf{k}}$ and ${\bf{k'}}$ data 
sets reveals that the ratio of intensities of satellites based on the 
same $H$ for 
${\bf{k}}$ and ${\bf{k'}}$ is different. 

\begin{figure}[hbt]
\caption {The $xy$-projection of the  ${\CuTeCl}$ ${\bf{k'}}$ magnetic structure with the spin tetrahedra at $-z$ and $z$.}
\includegraphics[width=86mm,keepaspectratio=true]{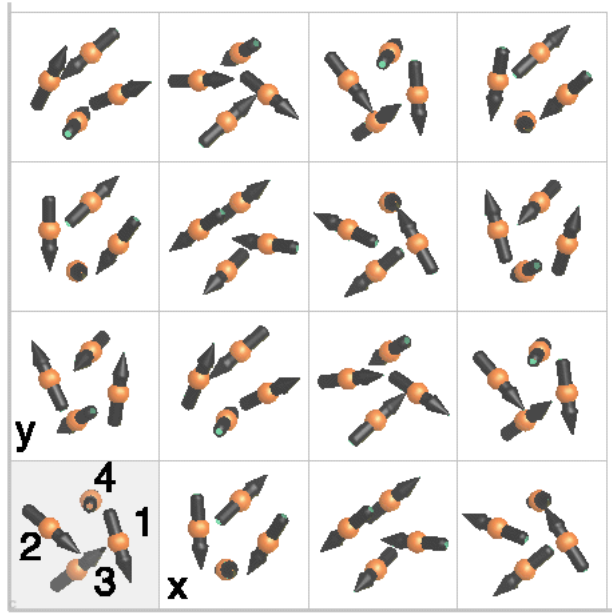}
\label{fig5}
            \end{figure}

In what follows we describe attempts to develop independent models for 
the ${\bf{k'}}$ structure from 59 reflections of the ILL
data set and for the ${\bf{k}}$  structure based on 18(25) 
magnetic reflections from the ILL
(TriCS) data sets.\\
It is worth noting that simple antiferromagnetic or ferromagnetic
arrangements are not compatible with the ICM wave vector. 
The lack of local symmetry at the 4(h) position and the
ICM wave vector implies that the magnetic moments of the four Cu${}^{2+}$ ions
in the cell are independent. 
The commensurate ${k_{z}} =\half$ means that corresponding spins in successive
unit cells along the $c$-axis are oppositely oriented.\\
In an ICM structure the moment ${\bf{S}_{jl}}$ of the $j$th 
ion in the $l$th unit cell can be expressed as
\begin{equation}
{ {\bf{S}_{jl}} = {\bf{A}_{j}} \cos({\bf{k\cdot
r_{l}}}+\psi_{j})+{\bf{B}_{j}} \sin({\bf{k\cdot r_{l}}}+\psi_{j}) }
\label{eq:eq4}
\end{equation}
The spin components are modulated by the propagation
vector ${\bf{k}}$. ${\bf{r_{l}}}$ is the radius vector to the origin of the
$l$th unit cell. ${\bf A_j}$ and ${\bf B_j}$ are orthogonal vectors which define
the magnitude and direction of the axes of the modulation on the $j$th atom,
whilst $\psi_{j}$ defines its phase. 
If $B$ (or $A$)=0, the structure is amplitude modulated:
the magnetic moment of each ion has a constant direction and varies sinusoidally from
cell to cell. Attempts to fit the data using such models led to unreasonably
high moments ($>$1 $\mu_B$/ion) at the wave maximum.\\
All other
situations describe more complex magnetic structures in which the spin
$j$ rotates from cell to cell in the plane defined by 
${\bf{A}_{j}}$ and ${\bf{B}_{j}}$. 
If ${|\bf{A}_{j}|} \ne {|\bf{B}_{j}|}$ the rotating spins
  $j$ have an elliptical envelope in the
$({\bf{A}_{j}},{\bf{B}_{j}})$ plane. When all the $({\bf{A}_{j}},{\bf{B}_{j}})$
planes are orthogonal or parallel to
the propagation vector, the specific structures are usually referred to as
helices or cycloids, respectively.\\
If ${|\bf{A}_{j}|}={|\bf{B}_{j}|}$ the 
envelope becomes circular and the atoms of site $j$ have the same moment 
in each cell.  We refer to this model with constant moments as the
generalized helix model.\\ 
A generalized helical spin arrangement is fully characterized by defining, for each 
independent atom, the amplitude 
$|A|=|B|$, two angles defining the plane in which the spins rotate and
the phase of the modulation. 
The angles chosen are the polar coordinates $\theta_j$, $\phi_j$ of
${\bf{B}_{j}}$ and the phase 
angle $\psi_j$(see 
Fig.~\ref{fig4}b). The vector ${\bf{A}_{j}}$ can always be chosen in the
$ab$ plane ($\theta_{\bf{A}_{j}}$=90); orthogonality of ${\bf{A}_{j}}$ and ${\bf{B}_{j}}$
is ensured by making
$\phi_{\bf{A}_{j}}$=$\phi_{\bf{B}_{j}}+90$. $\psi_j$ is the angle between
${\bf{S}_{j0}}$ and ${\bf{A}_{j}}$.
This description requires 12 independent parameters within the physically reasonable assumption that all the Cu${}^{2+}$ moments are equal.\\
A simulated annealing algorithm which provides a general 
purpose optimization technique to resolve
this kind of large combinatorial problem\cite{Kirkpatrick83, Juan93} was used to
generate possible models which  could subsequently be refined by a least squares 
procedure.\cite{CCSL}\\
In one class of models, collinear arrangements of  the Cu spins within 
a tetrahedron was imposed. This may results in either a tetramer with $S_{tet}$=0 
(${{J}_1}\gg {{J}_2}$) or  two
dimers with $S_{dim}$=0 (${{J}_1}\ll {{J}_2}$). 
These models would assume that the dominant coupling is antiferromagnetic. Such AF exchange
is poorly justified since: in the first case the ${{J}_1}$ coupling is
associated mainly with the Cu-O-Cu path, with an angle
Cu-O-Cu approximately 110 deg. In the second case, the path associated with the ${{J}_2}$
coupling is even more complex,  possibly involving the halogen 
orbitals. We therefore also tried collinear arrangements which
could arise from FM or partial FM exchange. None of the collinear arrangements 
give a reasonable fit to the experimental observations and models based on
tetramers or dimers can therefore be discarded.\\ 
Much better fits to the data
were obtained for models in which
the 4 Cu$^{2+}$ moments in each tetrahedron form two canted
pairs: Cu1-Cu2 and Cu3-Cu4. The pairs share a common $({\bf{A}_{j}},{\bf{B}_{j}})$ plane 
but the associated helices have different phases $\psi_j$ (see Table~\ref{tab1}). The difference between the phases 
defines the canting angle between spins of the pair $\alpha$ and this angle is the same for 
all tetrahedra in the structure. Using this model (Fig.~\ref{fig5}) and the propagation vector ${\bf{k'}}$ 
we obtained $\alpha_{12}$=38(6) deg, $\alpha_{34}$=111(14) deg and  $m_{12}$=1.27(6) $\mu_B$/pair, $m_{34}$=0.76(14) $\mu_B$/pair. The refined moment value of the Cu${}^{2+}$ ions is 0.67(1)$\mu_B$/ion and the magnetic reliability factor R${}_M$ is 10.7\% for this model.\\
Surprisingly, the above model does not give a good 
fit to the intensities of reflections indexed with the ${\bf{k}}$
wave vector; the relative arrangement of the pairs in the ${\bf{k}}$ and ${\bf{k'}}$ structures must be different. Unfortunately the ${\bf{k}}$ data sets are too limited to
allow a final model for this wave vector to be proposed. But these findings indicate that a number of ground states with equal or close energies might exist.\\
The complexity of the  ${\bf{k'}}$ magnetic structure implies 
a delicate interplay involving geometrical frustration, between the different
couplings within the
tetrahedra and the quite significant inter-tetrahedral interaction.
These together with the antisymmetric Dzyaloshinskii-Moriya
anisotropy interactions define the final spin arrangement. The model 
can account for the almost isotropic magnetic 
susceptibility\cite{Prester04}, and gives support to the
coexistence of AF and 90-deg couplings, that was  first proposed
from Raman experiments.\cite{Jensen03} The presence of the `canted 
pair' motif might suggest that the ${{J}_2}$ interaction prevails in 
the competition between all the above mentioned interactions. However,  
this question can be left open for future study.\\
The spin arrangement in $\CuTeBr$ cannot be determined from the available NPD data.
The magnetic moment derived, assuming the same magnetic structure as for the 
Cl-compound, is only 0.51(5)$\mu_B$/Cu${}^{2+}$ at 1.5~K, much less than in $\CuTeCl$.  
This low value may indicate that the Br-compound is 
closer to the quantum critical point than the 
Cl-analogue.\cite{Lemmens01, Gros03, Prester04} However, the similarity 
of the magnetic wave vectors does not guarantee the same magnetic structure 
for the two systems. Single crystal experiments with a high flux neutron 
diffractometer are necessary to determine the spin arrangement in the Br-compound.\\
We conclude that our neutron powder and single crystal diffraction 
experiments confirm the existence of long-range magnetic ordering in 
${\CuTeX}$ 
compounds below $T_{N}$. The magnetic order is propagated with the 
incommensurate wave vectors ${\bf{k}_{Cl}}\approx[0.150,0.422,$\half$]$ and 
${\bf{k}_{Br}}\approx[0.158,0.354,$\half$]$. The model proposed for 
the Cl-compound implies a canting of the spins and the presence of two canted pairs within the tetrahedra. 
New theoretical studies are needed to quantify the interplay 
between spin frustration and quantum fluctuations in these systems.\\
The work was performed at SINQ, Paul Scherrer Institute, Villigen, 
Switzerland and ILL reactor, Grenoble, France.
We thank Drs. M. Prester, H. Ronnow and Profs. A. Furrer, F. Mila for 
fruitful discussions and Drs. L. Keller, V. Pomjakushin for 
experimental assistance. The sample preparation was supported by the 
NCCR research pool MaNEP of the Swiss NSF.
\begin{table}
\caption{
The magnetic parameters from neutron diffraction studies of 
${\CuTeCl}$.
\label{tab1}}
\begin{ruledtabular}
\begin{tabular}{lclclclc}
\multicolumn{2}{c}{${\bf{k}}$}&\multicolumn{3}{c}{0.1505(8), 
0.4220(2), $\half$}&\multicolumn{3}{c}{0.67(1),$\mu_B/Cu{}^{2+}$}\\
\multicolumn{2}{c}{}&\multicolumn{2}{c}{$\theta_{\bf{B}}$}&\multicolumn{2}{c}{$\phi_{\bf{B}}$}&\multicolumn{2}{c}{$\psi$, deg}\\
\multicolumn{2}{c}{Cu1}&\multicolumn{2}{c}{57(9)}&\multicolumn{2}{c}{17(8)}&\multicolumn{2}{c}{0}\\
\multicolumn{2}{c}{Cu2}&\multicolumn{2}{c}{57(9)}&\multicolumn{2}{c}{17(8)}&\multicolumn{2}{c}{38(6)}\\
\multicolumn{2}{c}{Cu3}&\multicolumn{2}{c}{139(8)}&\multicolumn{2}{c}{-49(8)}&\multicolumn{2}{c}{151(10)}\\
\multicolumn{2}{c}{Cu4}&\multicolumn{2}{c}{139(8)}&\multicolumn{2}{c}{-49(8)}&\multicolumn{2}{c}{262(10)}\\
\end{tabular}
\end{ruledtabular}
\end{table}

\end{document}